\documentclass[acmtog, nonacm]{acmart}

\usepackage{booktabs} % For formal tables

\usepackage[ruled]{algorithm2e} % For algorithms

\SetAlFnt{\small}
\SetAlCapFnt{\small}
\SetAlCapNameFnt{\small}
\SetAlCapHSkip{0pt}

\usepackage{amsbsy}
\usepackage[mathscr]{euscript}
\usepackage{amsmath}
\usepackage{stackrel}
\usepackage{nicefrac}
\usepackage{siunitx}
\usepackage{soul}
\usepackage{xcolor}

\begin{document}
% Title portion
\title{Multisource Holography}

\author{Grace Kuo}
\affiliation{%
    \institution{Reality Labs Research, Meta}
    \country{USA}
}

\author{Florian Schiffers}
\affiliation{%
    \institution{Reality Labs Research, Meta}
    \country{USA}
}

\author{Douglas Lanman}
\affiliation{%
    \institution{Reality Labs Research, Meta}
    \country{USA}
}

\author{Oliver Cossairt}
\affiliation{%
    \institution{Reality Labs Research, Meta}
    \country{USA}
}

\author{Nathan Matsuda}
\affiliation{%
    \institution{Reality Labs Research, Meta}
    \country{USA}
}

\renewcommand\shortauthors{Kuo, G. et al.}

%%%%%%%%%%%%%%%%%%%%%%%%

\begin{abstract}

Holographic displays promise several benefits including high quality 3D imagery, accurate accommodation cues, and compact form-factors. However, holography relies on coherent illumination which can create undesirable speckle noise in the final image. Although smooth phase holograms can be speckle-free, their non-uniform eyebox makes them impractical, and speckle mitigation with partially coherent sources also reduces resolution. Averaging sequential frames for speckle reduction requires high speed modulators and consumes temporal bandwidth that may be needed elsewhere in the system.

In this work, we propose multisource holography, a novel architecture that uses an array of sources to suppress speckle in a single frame without sacrificing resolution. By using two spatial light modulators, arranged sequentially, each source in the array can be controlled almost independently to create a version of the target content with different speckle. Speckle is then suppressed when the contributions from the multiple sources are averaged at the image plane. We introduce an algorithm to calculate multisource holograms, analyze the design space, and demonstrate up to a 10 dB increase in peak signal-to-noise ratio compared to an equivalent single source system. Finally, we validate the concept with a benchtop experimental prototype by producing both 2D images and focal stacks with natural defocus cues.

\end{abstract}

\begin{teaserfigure}  
\vspace{3mm}
  \includegraphics[width=\linewidth]{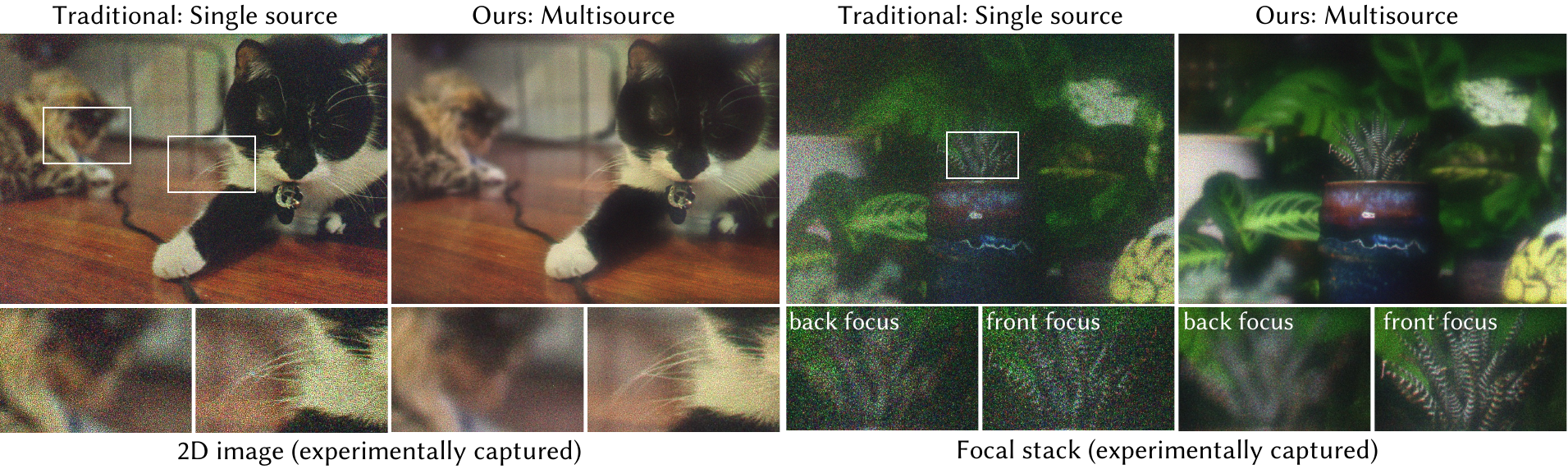}
  \caption{\label{fig:teaser}
  We propose a new architecture for holographic displays specifically designed for speckle reduction. Instead of a single coherent source of illumination, our design uses a grid of multiple sources, which sum incoherently at the image plane. By using two spatial light modulators (SLMs) with an air gap in between, we break correlations between the multiple sources enabling high resolution holograms with significantly suppressed speckle. We experimentally demonstrate speckle reduction on both 2D images (left) and focal stacks with natural defocus blur (right).
  }
\vspace{3mm}
\end{teaserfigure}

\maketitle

%%%%%%%%%%%%%%%%%%%%%%%%%%%%%%%%%%%%%%%%%%%%%%%%%%%%%%%%%%%%%%%%%%%%%%%%%%%%%%%%%%%%%%%%

\section{Introduction}
\label{sec:intro}

Computer generated holography uses a spatial light modulator (SLM) to mimic the wavefront coming from a three-dimensional (3D) object. This enables high resolution displays with accurate per-pixel focal cues, and recent user studies demonstrated that holographic displays have the potential to drive the human accommodation response~\cite{kim2022accommodative}, offering a solution to the vergence-accommodation conflict of stereoscopic displays~\cite{hoffman2008vergence}. Holography is a particularly promising technology for head-mounted displays (HMDs) since it  also enables compact form-factors, can compensate for optical aberrations, and can correct for eyeglass prescriptions entirely in software~\cite{maimone2017holographic}.

However, holographic displays rely on spatially coherent illumination to achieve 3D cues~\cite{lee2020light}, which can create speckle in the displayed content. Speckle is a phenomenon that occurs with coherent light when random path length differences interfere at the image plane, creating a noisy pattern of dark and bright spots due to random constructive and destructive interference \cite{goodman2007speckle}. This effect is undesirable since it hides details in the hologram and creates noisy, visually unappealing images. Although reducing the illumination coherence can suppress speckle, it also reduces resolution and depth of field~\cite{deng2017coherence}.

Smooth phase holograms offer a different option to control speckle: by removing randomness in the image plane phase, all interference is constructive and speckle is eliminated~\cite{maimone2017holographic}. However, these holograms have highly non-uniform energy distribution in the eyebox, greatly reducing practicality~\cite{yoo2021optimization}. In addition, the focal cues generated are limited and often exhibit unnatural ringing. In user studies, these focal cues were not effective at driving accommodation ~\cite{kim2022accommodative}.

Another option for speckle reduction is temporal multiplexing, when several frames with unique speckle patterns are shown in rapid sequence such that they are visually averaged by the eye. However, this requires high speed SLMs with frame rates in the kilohertz range, and the amount of despeckling increases sub-linearly with the number of averaged frames. Reducing the number of frames needed for speckle control could increase the amount of speckle reduction with the same number of frames, could allow for more flexibility choosing modulators, or could free temporal bandwidth to address other challenges, for example, increasing eyebox size or field of view (FoV) through scanning. 

We propose a novel architecture for speckle reduction in holographic displays that can create natural defocus cues with a uniform eyebox at the full SLM resolution, all in a single frame. To do this, we modify the illumination setup of a traditional holographic display, which typically consists of a single light source that  generates a coherent plane wave at the SLM. In our architecture, we replace this single source with a grid of multiple sources, which each generate a plane wave at a different angle of incidence. By using sources that are incoherent with each other, the speckle patterns from each source average at the image plane, reducing speckle contrast. 

However, with a single SLM, each source creates a shifted copy of the same hologram, creating haze and doubling in the displayed content. To address this, we propose using two SLMs spaced a few millimeters apart axially. This arrangement creates a modulation response that varies with the angle of incidence, similar to how volume holograms use their thickness to create angular selectivity ~\cite{heanue1994volume}. With the two SLMs, we can independently control the output image from each source, removing the doubling artifacts while continuing to get the speckle-reduction benefits of the multiple sources. We refer to this architecture, including both the array of sources and the two SLMs, as \textit{multisource holography}.

In summary, we make the following contributions:
\begin{itemize}
    \item We introduce the multisource holography architecture and a corresponding hologram generation algorithm. We demonstrate full resolution holograms with natural defocus cues and uniform eyebox in a single frame.
    \item In simulation, we demonstrate improvements of 10 dB in peak signal-to-noise ratio (PSNR) compared to an equivalent single source system with the same degrees of freedom. We further show that multisource holography with no temporal averaging outperforms temporal multiplexing with 6 jointly optimized frames.
    \item We analyze how the source spacing and number of sources impact hologram quality and provide guidance on the minimum spacing needed to achieve full resolution.
    \item We validate the multisource holography concept with a full-color benchtop prototype. We introduce a customized calibration procedure and experimentally demonstrate low speckle holograms for both planar images and focal stacks with natural blur.
\end{itemize}

\section{Related Work}
\label{sec:related}

\paragraph{Smooth Phase Holograms}

As described above, smooth phase holograms eliminate speckle by enforcing near-constant phase at the image plane, which removes randomness and ensures interference between neighboring points is always constructive. Enforcing a specific phase at the image plane requires complex modulation at the SLM, so a practical option is the double phase amplitude coding (DPAC) method, which can almost entirely remove speckle~\cite{maimone2017holographic, shi2021towards}.  Even without complex modulation, one can achieve low phase variation at the image plane through gradient descent with uniform phase initialization at the SLM \cite{chakravarthula2019wirtinger} or by explicitly enforcing a piece-wise constant phase in the loss function~\cite{Choi2021NeuralDisplays}. Although these smooth phase approaches can create high quality and speckle-free two dimensional (2D) images, defocus blur is limited and contains unnatural ringing. To address this issue, holograms can be optimized to target natural-appearing blur while encouraging image phase to remain smooth \cite{kavakli2023realistic, yang2022diffraction}. However, the amount of blur is still limited, and all smooth phase holograms concentrate energy into a small region of the eyebox, making these systems very sensitive to eye movement and imperfections in the user's eye. Furthermore, in a recent user study, smooth phase holograms were not effective at driving accommodation \cite{kim2022accommodative}.

\paragraph{Random Phase Holograms}

Random phase at the image plane, which generates scattering similar to a diffuse object, enables natural defocus blur and uniform energy distribution in the eyebox~\cite{yoo2021optimization}. However, this same randomness reintroduces speckle due to interference from different random path lengths. For 2D images, by letting the phase at the image plane be a free variable, one can use iterative approaches to shape the phase such that speckle is minimized from a particular viewing angle \cite{Saxton1972GerchbergSaxtonBeams, fienup1982phase}. Adding spatial ``don't care'' regions can further enhance image quality in the regions of interest \cite{georgiou2008aspects}. However, for 3D content, the number of degrees of freedom on the SLM is insufficient to suppress speckle everywhere at once. One option is to let out-of-focus content be unconstrained, which enables better in-focus imagery but creates additional speckle in defocused regions \cite{Choi2021NeuralDisplays, kuo2020high}. Generating natural defocus blur with low speckle over the whole volume requires additional despeckling approaches that cannot be achieved through the algorithm alone.

\paragraph{Partial Coherence}

Decreasing the coherence of the illumination can reduce speckle by imposing an incoherent blur on the output image through wavelength diversity (temporal partial coherence) or angular diversity (spatial partial coherence)~\cite{deng2017coherence, zhao2022speckle}. Spatial partial coherence in holographic displays has been demonstrated using an echelon stair~\cite{lee2019speckle}, and temporal partial coherence has been demonstrated with different light sources, such as light emitting diodes (LEDs) \cite{moon2014holographic, kozacki2022led} and superluminescent LEDs (SLEDs)~\cite{primerov2019p,peng2021speckle}. However, partially coherent sources result in a direct trade-off between resolution and speckle reduction, which is incompatible with a high resolution, low speckle display. \citet{lee2020light} designed a partially coherent light source specifically to balance resolution, depth of field, and speckle, but the trade-off still exists so despeckling is limited without further resolution reduction. Similar coherence properties have been explored in the context of interferometric 3D sensing~\cite{kotwal2023swept} and transmission matrix characterization~\cite{gkioulekas2015micron, kotwal2020interferometric}, and we refer interested readers to these sources for an in-depth analysis.

\paragraph{Temporal Multiplexing}

To achieve despeckling without sacrificing resolution, one can display many holograms in sequence, each with a unique speckle pattern. Due to human persistence of vision, the user sees an average of the displayed images, effectively suppressing speckle. 
Systems with 8 to 24 frames of temporal multiplexing per color have been demonstrated with high speed modulators such as digital micromirror devices (DMDs) \cite{lee2020wide, curtis2021dcgh}, ferro-electronic liquid crystal on silicon (FLCoS) \cite{lee2022high}, and micro-electromechanical systems (MEMS) \cite{choi2022time}. These prior works achieve state-of-the-art image quality for temporal multiplexing by jointly optimizing all frames and accounting for the limited bit depth of these high speed SLMs. However, to create a fully life-like HMD, one needs to refresh content at least 1.8 kHz \cite{cuervo2018creating}; achieving these refresh speeds with temporal multiplexing requires updating content between sub-frames, which current algorithms do not support.

In addition, reducing the number of frames needed for speckle control could free the temporal bandwidth for other uses. For example, \citet{lee2020wide} demonstrated increased viewing angle (in other words, increased \'etendue) by scanning illumination angle over time. Approaches like these could help overcome the fundamental \'etendue limits~\cite{park2022holographic} of holographic displays, but they  reduce the amount of temporal bandwidth available for despeckling.

\paragraph{Multiple Modulators}

Our system is capable of reducing speckle in a single frame through the use of two cascaded SLMs, taking advantage of the compression that layered displays can provide. In conventional optics, layered modulators can be used to break the trade-off between spatial and angular resolution in light field displays~\cite{wetzstein2012tensor}. Similarly, in diffractive optics, \citet{ye2014toward} showed that static layered diffractive elements can control the 4D bidirectional reflectance distribution function (BRDF) under incoherent illumination, and \citet{peng2017mix} used pairs of static diffractive optical elements (DOEs) to create different holograms based on the relative translations between the DOEs. Although these prior publications have different application spaces, they demonstrate that two layered modulators can create several (more than two) different images, highlighting the compressive nature of the layered displays. 

In our system, we take advantage of compression in layered displays to achieve more despeckling than in non-compressive systems (such as temporal multiplexing) with the same number of degrees of freedom. We note that interferometer-inspired setups \cite{choi2021optimizing, wang2019holographic}, which also use multiple SLMs for image quality enhancement, are not designed to take advantage of potential compression, and therefore have limited speckle reduction based on the degrees of freedom in the two modulators.

\paragraph{Multiple Incoherent Sources}

Despeckling in our system is achieved through multiple discrete sources of illumination that are incoherent with each other. To our knowledge, the only prior work with similar illumination is that of \citet{jo2022multi} in which multiple sources are used for \'etendue expansion while simultaneously providing some despeckling. Like our work, they show that multiple sources with a single modulation plane create uncontrollable replicas in the final image. However, they use a binary amplitude mask in the Fourier plane to break correlations between replicas, as where we use a second SLM with a small air gap, which is more amenable to a compact system and provides additional degrees of freedom for better image quality. Finally, since \citet{jo2022multi} target \'etendue expansion as their application, they fix the number and locations of the sources such that, at any position in the image, a maximum of 9 different sources are averaged for speckle reduction. We demonstrate that speckle reduction can be dramatically increased with more sources, and we analyze the effect of number of sources and source spacing on image quality.

\paragraph{Camera-Based Calibration}

Even if speckle is theoretically reduced, any non-idealities in the optical system can cause additional speckle in practice due to mismatch between the model used in optimization and the true system. To account for imperfections, one can design a model of the optical system with learnable parameters, then fit the unknowns in an offline calibration process using experimentally captured data \cite{peng2020neural, Choi2021NeuralDisplays, Chakravarthula2020LearnedDisplays, kavakli2022learned}. A special case of camera-based calibration is the ``active'' approach proposed by \citet{peng2020neural}, in which the SLM pattern is fine-tuned online to a particular image based on camera feedback. Although these holograms do not generalize to new content, they highlight what is feasible with a given experimental system. To best demonstrate the potential of multisource holography, we use both offline calibration with a physically-based model and online active camera-in-the-loop.

\begin{figure*}
    \centering
    \includegraphics[width=\linewidth]{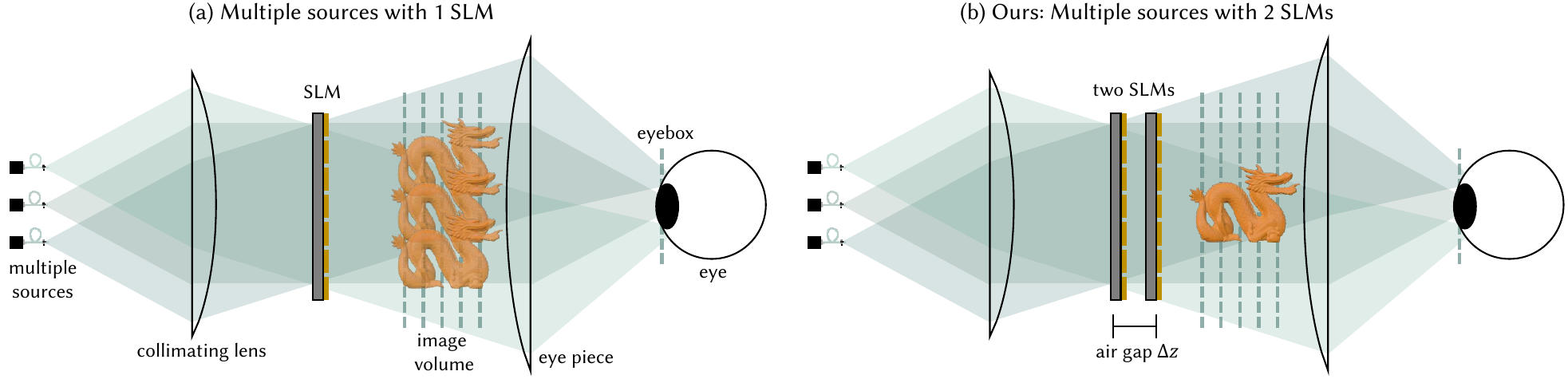}
    \caption{ \textbf{System Architecture:} 
    Multisource holography uses an array of mutually incoherent sources that each generate a plane wave at a different angle. (a)~With a single SLM, all sources are modulated with the same pattern but propagate in different directions, creating replicas of the content. Generating an image with this configuration is a poorly posed problem. (b) We propose adding a second SLM a small distance $\Delta z$ in front of the first. This enables different modulation function for different angles of incidence, enabling unique content for each source. By jointly optimizing the two SLM patterns, the  holograms from each source line up correctly, removing replica artifacts. Since the sources are incoherent with each other, their intensities add at the image plane which suppresses speckle through averaging.
}
    \label{fig:diagram_simple}
\end{figure*}

\section{System Overview}
\label{sec:methods}

A traditional holographic display uses an SLM to shape an incoming coherent beam to form an image. Denoting the complex modulation function of the SLM as $\mathbf{s}(\vec{x})$, we can write the image formation model as
\begin{equation}
\begin{aligned}
    \mathbf{g}_z(\vec{x}) &= \mathcal{P}_{z} \big\{ \mathbf{p}(\vec{x}) \odot \mathbf{s}(\vec{x}) \big\}, \\
    \mathbf{I}_z(\vec{x}) &=   | \mathbf{g}(\vec{x}) |^2,
    \label{eq:singlesource_forward_model}
\end{aligned}
\end{equation}
where $\vec{x}$ is the 2D spatial coordinate at the SLM, $\mathbf{g}_z (\cdot)$ and $\mathbf{I}_z (\cdot)$ are, respectively, the electric field and intensity a distance $z$ from the SLM, and $\mathbf{p}(\cdot)$ is the complex field illuminating the SLM, which is most commonly a plane wave of unit energy, $\mathbf{p}(\vec{x}) = 1$. Finally, $\odot$ denotes pointwise multiplication, and $\mathcal{P}_{z} \{\cdot\}$ is the angular spectrum method (ASM) propagation operator defined as
\begin{align}
\mathcal{P}_z\left\{\mathbf{s}(\vec{x})\right\} &=  \mathcal{F}^{-1} \bigl\{ \mathcal{F}\{\mathbf{s}(\vec{x}) \}  \odot \mathcal{H}_z(\vec{u} ) \bigr\} ,
\label{eq:ASM_propagation} \\
\mathcal{H}_z( \vec{u} ) &= \begin{cases} \exp \left( \frac{2 \pi j z}{\lambda} \sqrt{1- \|\lambda \vec{u} \|^2  } \right), & \text { if } \sqrt{\|\vec{u}\|^{2}}<\frac{1}{\lambda}, \\
0, & \text { otherwise, }\end{cases}
\label{eq:ASM_kernel}
\end{align}
where $\mathcal{F}\{\cdot\}$ is the 2D Fourier transform operator and $\vec{u}$ is the 2D coordinate in frequency space~\cite{matsushima2009band}. Here, we assume monochromatic illumination with wavelength $\lambda$; see Supplement for an extension to broadband sources.

To generate a hologram, one can use first order methods like gradient descent to optimize for an SLM pattern that creates a given target image:
\begin{align}
    \mathbf{s} = \operatorname*{argmin}_\mathbf{s} \sum_z \| \mathbf{I}_z(\vec{x}) -  \mathbf{\hat{I}}_z(\vec{x}) \|_2^2,
    \label{eq:optimization_single_source}
\end{align}
where $\mathbf{\hat{I}}_z(\vec{x})$ is the target intensity at a given plane, and optimization is performed over a dense range of propagation distances in the volume of interest. 

To encourage natural defocus cues, we render the target images with realistic blur based on incoherent illumination where the blur kernel size is determined by the maximum diffraction angle of the SLM (see Supplement for details). However, the holographic display aims to control a 3D volume of light using only a single 2D SLM pattern, making the optimization problem overdetermined. As a result, the 3D volume cannot be matched exactly and uncontrollable speckle noise is visible in the image, particularly as the image volume grows.

\subsection{Despeckling with Multiple Illumination Sources} \label{sec:multisource_1slm}

Our goal is to reduce speckle in holographic displays. Our basic strategy is common in the despeckling literature:
produce several versions of the image that each have a unique speckle pattern, and when these copies of the image are superimposed, the speckle is reduced through averaging~\cite{goodman2007speckle}. To create different versions of the image, we propose using multiple sources of illumination. When sources are placed at different locations behind a collimating lens, as shown in Fig.~\ref{fig:diagram_simple}a, each source illuminates the SLM from a different angle:
\begin{align}
    \mathbf{p}(\vec{x}; \vec{m_i}) = e^{j (\vec{x}\cdot\vec{m_i})},
    \label{eq:angle_of_incidence}
\end{align}
where $\cdot$ denotes inner product and $\vec{m_i}$ is the phase slope (in radians per meter) of the $i$-th source at the SLM plane, which is related to illumination angle of incidence by 
\begin{align}
    \vec{\theta} = \frac{\lambda\vec{m}_i}{2\pi}.
\end{align}
Note that, unlike the work of \citet{jo2022multi}, we are not using the sources to expand \'etendue. Therefore, we choose small slopes for $\vec{m}_i$, within the range of angles that the SLM is able to create natively.

If the different sources are all incoherent with each other, they will not exhibit interference effects at the image plane when they are combined.
Instead, the multisource image is the sum of the individual source intensities as follows:
\begin{align}
    \mathbf{g}_{z, m_i}(\vec{x}) &=  \mathcal{P}_{z} \big\{ \mathbf{p}(\vec{x}, \vec{m}_i) \odot \mathbf{s}(\vec{x}) \big\} \\
    \mathbf{I}_{z}(\vec{x}) &= \sum_i \big| \mathbf{g}_{z, m_i}(\vec{x}) \big|^2
    \label{eq:incoherent_sum}
\end{align}

This achieves part of our goal: each source creates a unique speckle pattern, so speckle contrast is reduced when the individual source intensities are combined. However, for a useful display, the final output intensity $\mathbf{I}_{z}$ should have the potential to be shaped into arbitrary target images, which is not the case in this configuration. To demonstrate the problem, we make the small angle approximation to the ASM kernel, and derive the following relationship (see Supplement for derivation):
\begin{align}
    \mathbf{g}_{z, m_i}(\vec{x}) =  \mathbf{g}_{z,0}(\vec{x} + \tfrac{ \lambda z}{2 \pi} \vec{m}_i)  \odot e^{j  (\vec{x} \cdot \vec{m}_i)}
\label{eq:memory_effect}
\end{align}
where $\mathbf{g}_{z,0}(\cdot)$ is the electric field from an on-axis plane wave ($\vec{m}=\vec{0}$).
This means that the output electric field from the $i$-th source is a translated copy of electric field with on-axis illumination, up to a carrier wave. In other words, a single ideal SLM has infinite memory effect~\cite{freund1988memory}. 

Based on Eqs.~\ref{eq:incoherent_sum} and \ref{eq:memory_effect}, the total intensity with all the sources can be written as
\begin{align}
    \mathbf{I}_{z}(\vec{x}) = \big| \mathbf{g}_{z, 0}(\vec{x}) \big|^2 * \sum_i \delta(\vec{x} + \tfrac{ \lambda z}{2 \pi} \vec{m}_i),
    \label{eq:deconvolution_problem}
\end{align}
where $*$ denotes a 2D convolution.
Therefore, producing a given output image $\mathbf{I}_{z}$ requires deconvolving the source locations. This is a very poorly posed problem, regardless of the $\vec{m}_i$ used, since the result of the deconvolution, $\big| \mathbf{g}_{z, 0}(\vec{x}) \big|^2$, is a physical quantity that must be nonnegative. As a result, multiple sources with a single SLM is not a viable solution for holographic displays.

\begin{figure*}
    \centering
    \includegraphics[width=\linewidth]{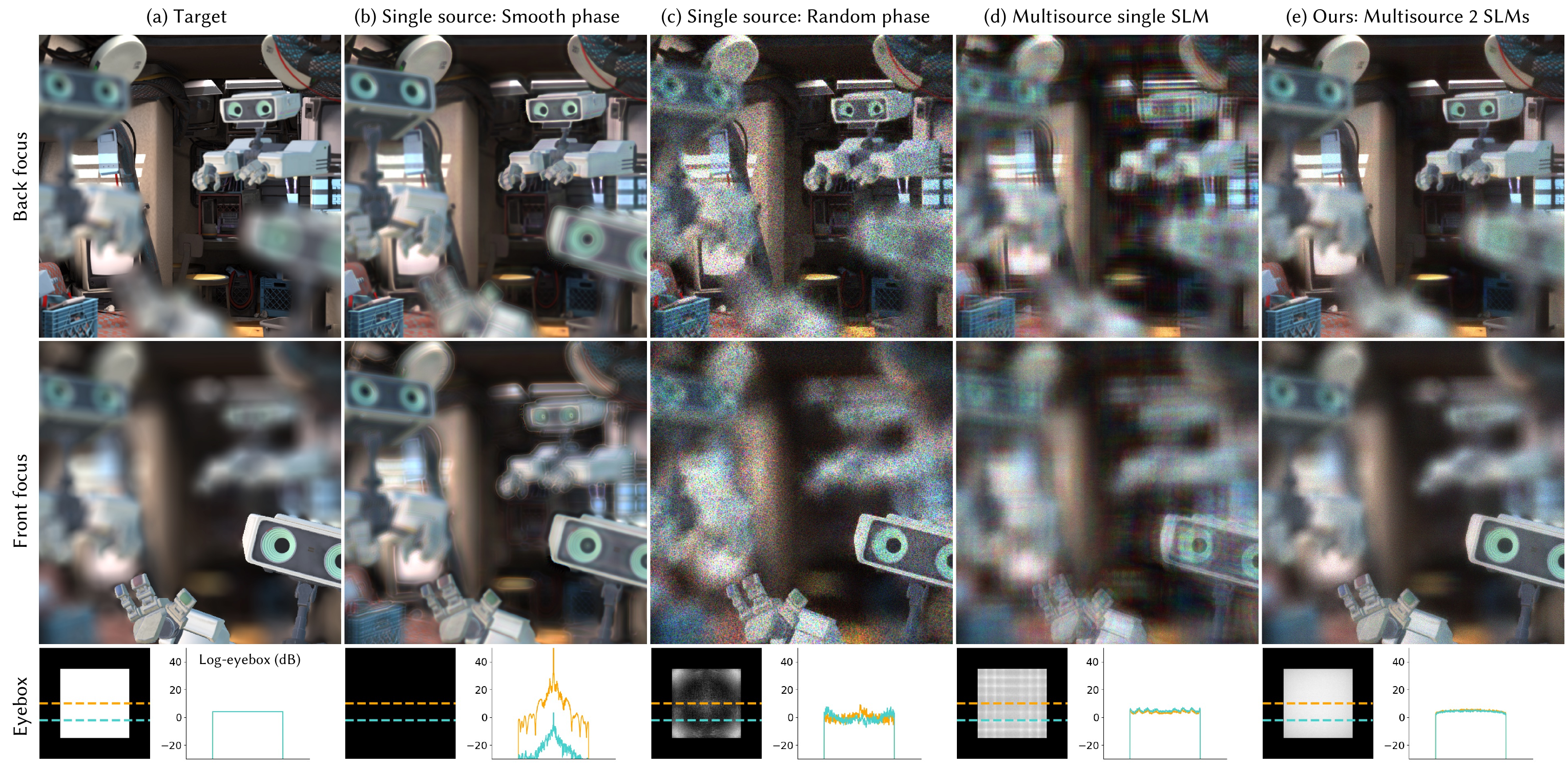}
    \caption{ \textbf{Single Source vs. Multisource (Simulation):}
    In simulation, we compare four methods for generating a target focal stack (a) with natural defocus cues. (b) A traditional single source hologram optimized with smooth phase has no speckle but there are ringing artifacts in the defocused regions. More importantly, the energy distribution in the eyebox (bottom row) is extremely non-uniform (note that plots are logarithmic) with a large peak in the center, which makes the display sensitive to eye imperfections and requires precise, low latency eye tracking and 3D pupil steering for a usable display. (c) A single source hologram with random phase achieves an approximately uniform eyebox distribution, but the image is corrupted by severe speckle. (d) Multiple sources reduce speckle, but with a single SLM, correlations between the outputs of each source create haze and doubling in the displayed hologram. (e) Our multisource holography approach uses two SLMs (here, one phase SLM and one amplitude SLM) to break correlations between the individual source outputs. This removes the low frequency artifacts in (d) while preserving the speckle reduction.  Although (e) uses two SLMs, all simulations have the same degrees of freedom since (b)-(d) are simulated with a single complex SLM. Out of these approaches, only multisource holography is capable of creating high quality focal stacks with a practical energy distribution in the eyebox. 
}
    \label{fig:concept}
\end{figure*}

\subsection{Multisource Holography with Two Modulators} \label{sec:multisource_2SLMs}

In order to display arbitrary content with multiple incoherent sources, there cannot be strong correlations between the different source outputs. We want each angle of illumination to generate a unique pattern, and this requires that the modulator have an angularly selective response. We achieve this requirement by adding a second SLM a distance $\Delta z$ from the first, as shown in Fig. \ref{fig:diagram_simple}b, which yields the following image formation model:
\begin{align}
    \mathbf{I}_{z}(\vec{x}) = \sum_i \bigg|\mathcal{P}_{z} \Big\{ \mathcal{P}_{\Delta z} \big\{ \mathbf{p}(\vec{x}; \vec{m}_i) \odot \mathbf{s}_1(\vec{x}) \big\}  \odot \mathbf{s}_2(\vec{x}) \Big\}\bigg|^2,
    \label{eq:multisource_forward_model}
\end{align}
where $\mathbf{s}_1(\cdot)$ and $\mathbf{s}_2(\cdot)$ are the modulation functions of the two SLMs. To see how the second SLM breaks the correlation between sources, let $\mathbf{g}_{\Delta z, 0} (\vec{x})$ be the electric field just before the second SLM given on-axis illumination. Applying Eq.~\ref{eq:memory_effect}, we can describe the electric field after the second SLM as
\begin{align}
    \mathbf{g}_{\Delta z, 0} (\vec{x} + \tfrac{ \lambda \Delta z}{2 \pi} \vec{m}_i) \odot \mathbf{s}_2(\vec{x}) \odot e^{ j  (\vec{x} \cdot \vec{m}_i)}.
\end{align}
Here, the electric field is translated based on the source angle, then pointwise multiplied by the modulation function of the second SLM. As long as the relative translation between any two sources is at least one SLM pixel, then the output fields (and therefore the final intensities) will be substantially decorrelated, breaking the memory effect~\cite{freund1988memory}. This gives the following condition on the source spacing:
\begin{align}
\Delta m \geq \frac{2 \pi p}{\lambda \Delta z},
\label{eq:memory_effect_condition}
\end{align}
where $\Delta m$ is the spacing between sources ($\Delta m = \|\vec{m}_i - \vec{m}_j\|^2$), and $p$ is the SLM pixel pitch, assumed be the same for both SLMs.

As long as Eq.~\ref{eq:memory_effect_condition} is met, our multisource holography setup can create different content for each source. Conceptually, each source ``sees'' a different relative translation between the two SLM patterns. Therefore, we would like to design the SLMs so  each of these translations creates the desired target image for each source. This is similar to the work of \citet{peng2017mix}, where pairs of static DOEs are combined with different translations to create unique images. Based on their results, where several unique holograms were created from a single DOE pair, we expect our system can also create the desired output for more than two sources simultaneously, even though there are only two SLMs. In other words, we expect the system to be compressive, which allows our system to generate more incoherent copies of the image, resulting in more despeckling, than other systems (for example, temporal multiplexing) with the same number of degrees of freedom.

In practice, we jointly solve for both SLM patterns using the using the model in Eq.~\ref{eq:multisource_forward_model} and solving the optimization problem,
\begin{align}
    \mathbf{s}_1, \mathbf{s}_2 = \operatorname*{argmin}_{\mathbf{s}_1, \mathbf{s}_2} \sum_z \| \mathbf{I}_z(\vec{x}) -  \mathbf{\hat{I}}_z(\vec{x}) \|_2^2,
    \label{eq:optimization_problem}
\end{align}
using ADAM \cite{kingma2014adam}.

\begin{figure*}
    \centering
    \includegraphics[width=\linewidth]{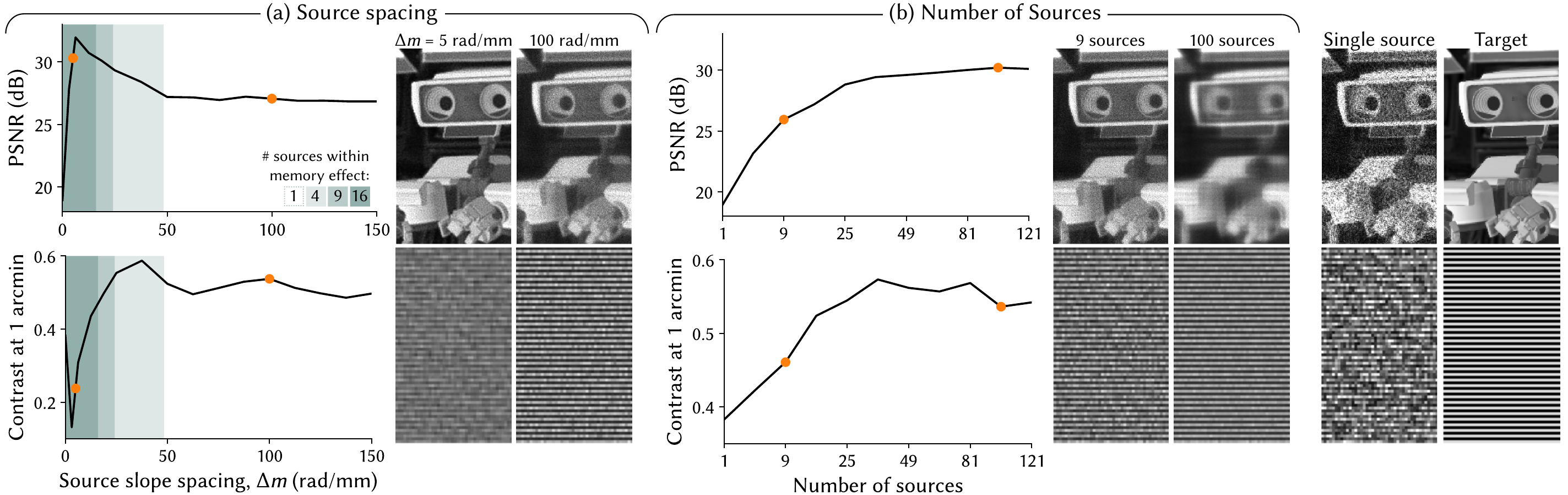}
    \caption{\textbf{Source Configuration Analysis:} We assess the impact of source spacing (a) and number of sources (b) on peak signal-to-noise-ratio (PSNR) for a natural scene (top) and contrast at 1 arcmin (bottom). Here, we assume the system is scaled so 1 arcmin corresponds to the maximum SLM resolution. (a)~When sources are close together, all sources are within the memory effect region (i.e. do not meet Eq.~\ref{eq:memory_effect_condition}), so each source generates a similar output, creating blur in the final image. Although a small blur increases PSNR, it decreases resolution creating a dip in the contrast metric at small spacings. As the source spacing increases, more sources leave the memory effect region and contrast at 1 arcmin increases, demonstrating that full resolution is possible when the sources are spaced sufficiently far apart. Example images at two different spacings (indicated by the orange dots) are shown on the right. (b)~As the number of sources grows, PSNR increases due to better speckle suppression. However, for large numbers of sources the SLMs cannot fully control the outputs of all sources, creating haze in the final image (see 100 source example). This effect is captured by the contrast metric, which decreases after about 36 sources.
    }
    \label{fig:analysis}
\end{figure*}

\section{Simulation} \label{sec:simulation}

To demonstrate the improvements that multisource holography provides, we optimize holograms in simulation to generate a focal stack with rendered incoherent blur, similar to what one would see in a natural scene (Fig. \ref{fig:concept}a). Our focal stack, shown in Fig.~\ref{fig:concept}a, covers a \SI{10}{\milli\metre} range in SLM space (from $z = \SI{15}{\milli\metre}$ to $z = \SI{25}{\milli\metre}$) with a blur radius of 4 pixels per millimeter of defocus, matched to the maximum diffraction angle of an SLM with  $\SI{8}{\micro\metre}$ pixels (see Supplement for an explanation of these parameters). Simulations are conducted in red-green-blue ($\lambda = \SI{640}{\nano\metre}, \SI{520}{\nano\metre}, \SI{450}{\nano\metre}$) assuming monochromatic illumination, and we supervise the loss at 15 evenly spaced planes.
Optimization is implemented in PyTorch on an Nvidia A6000 GPU at $2\times$ the SLM resolution in each direction to avoid aliasing. %(see Supplement for details).

\subsection{Single Source Holograms} \label{sec:sim_single_source}

Using the traditional configuration with a single source and single SLM, it's difficult to create practical, high quality holograms with natural defocus. To demonstrate the challenges, we solve Eq.~\ref{eq:optimization_single_source} using the model in Eq.~\ref{eq:singlesource_forward_model}, where we assume a complex SLM. Although most off-the-shelf modulators control either phase or amplitude, but not both, we choose a complex SLM for this simulation as it has the same number of degrees of freedom as our 2-SLM multisource approach.

For single source holograms, the SLM initialization has a big impact on the result. We consider two different initializations: constant (in both phase and amplitude) versus uniform random. In both cases, after initialization we iteratively optimize the SLM pattern using ADAM in Pytorch based on the loss function in Sec.~\ref{sec:methods}; however, even after optimization the final SLM pattern is influenced by the starting point. For example, with constant initialization, the phase of $\mathbf{g}_z(\vec{x})$ tends to be low variance~\cite{yoo2021optimization}, resulting in a smooth phase hologram (Fig.~\ref{fig:concept}b). Similarly, with random initialization, $\mathbf{g}_z(\vec{x})$ tends to be high variance, resulting in a random phase hologram (Fig.~\ref{fig:concept}c). As shown in Fig.~\ref{fig:concept}b, the smooth phase simulation has low speckle noise, but exhibits unnatural ringing in defocused regions. In contrast, the random phase hologram has more natural defocus effects but contains substantial speckle noise. 

Although ringing may seem like an acceptable trade-off for speckle removal, smooth phase holograms are impractical for near-eye displays due to their eyebox energy distribution, shown in the bottom row for the green channel. The eyebox, which is created by an eyepiece of focal length $f$, is the area where the user's pupil is located (see Fig. \ref{fig:diagram_simple}). The electric field at the eyebox, $\mathbf{e}(\cdot)$, is described by
\begin{align}
\mathbf{e}(\vec{u}) = \mathcal{F} \{\mathbf{g}_{z_0}(\vec{x}) \},
\end{align}
where $z_0$ is the propagation distance from the SLM to the focal plane of the eyepiece, and $\vec{u}$ is the spatial coordinate at the eyebox ($\vec{u} = \vec{x}/\lambda f$)~\cite{goodman2005introduction}.

The eyebox of the smooth phase hologram (Fig.~\ref{fig:concept}b, bottom) has a very strong peak in the center, with almost 5 orders of magnitude more energy in the peak than in the eyebox periphery. This presents several challenges for a practical display since the eyebox energy is mostly concentrated into an area only a handful of microns across. First, this means that small eye movements, even those contained within the theoretical eyebox~\cite{kuo2020high}, cause the eye to miss the peak, and then the user will not see the image. See Fig.~\ref{fig:pupil_invariance} for an example of this effect. Second, since the light is concentrated into a small point on the user's pupil, ``floaters'' (debris in the vitreous humor) or other imperfections in the eye can cause substantial artifacts in the hologram. These imperfections are barely noticeable in daily life since the image on the retina is typically an integral over the full pupil; however, in a smooth phase hologram, only a small part of the pupil is sampled. Computationally removing the effects of floating debris is unrealistic as it would require detailed, real-time mapping of every user's eyes. Even if eye imperfections could be overcome, user studies suggest that the small eyebox of smooth phase holograms cannot effectively drive accommodation: \citet{kim2022accommodative} found much lower accommodative gain for smooth phase holograms compared to random phase holograms. As a result of all of these restrictions, we believe smooth phase holograms cannot achieve compelling 3D content with good image quality for all users.

Random phase holograms, on the other hand, simulate a diffuse surface at the object, which scatters light to cover the full theoretical eyebox (Fig.~\ref{fig:concept}c, bottom), but this comes at the cost of speckle.
%However, the random path length differences caused by the random phase create speckle. 
Although random phase holograms can be low speckle for a 2D scene, for a 3D focal stack the degrees of freedom on the SLM are insufficient to control speckle at all planes, even with a complex modulator. Not only does this speckle hide detail and make images visually unappealing, the high frequency speckle can also interfere with the human accommodation response which expects low spatial frequencies in defocused regions \cite{kim2022accommodative}. As a result, with a single source, neither smooth nor random phase holograms can produce high quality images that drive accommodation without additional speckle reduction.

\subsection{Multisource Holograms} \label{sec:sim_multisource}

Our multisource holography approach achieves the benefits of random phase holograms while adding substantial despeckling to reduce noise and produce more natural defocus cues. However, as described in Sec. \ref{sec:multisource_1slm}, adding more sources with a single SLM results in a poorly posed optimization problem that is not able to display arbitrary content. Figure~\ref{fig:concept}d shows an example with a $4\times4$ grid of sources and a single complex SLM. Although there is substantial noise reduction compared to random phase with a single source, the resulting image contains low frequency artifacts, as expected, due to the strong correlations between the outputs of each source. However, even with a single SLM, the multisource hologram is able to create an approximately uniform eyebox when initialized with a random pattern, albeit with some periodic structure due to the sources.

Our final design uses an array of sources with two SLMs, as described in Sec.~\ref{sec:multisource_2SLMs}, where the gap between the SLMs creates an angularly selective response that breaks the correlations between sources. Figure \ref{fig:concept}e shows a simulation of this configuration with a $4\times4$ grid of sources such that all sources are outside the memory effect region (Eq.~\ref{eq:memory_effect_condition}) for all wavelengths of interest. Of our two SLMs, spaced $\Delta z = \SI{2}{\milli\metre}$ apart, the first SLM modulates phase only, and the second SLM modulates amplitude only, creating the exact same degrees of freedom as in the prior simulations. We initialize the SLMs with uniform random phase and amplitude, respectively. Figure \ref{fig:concept}e demonstrates that the second SLM successfully breaks the correlations between the sources, removing the low frequency artifacts of Fig.~\ref{fig:concept}d while substantially suppressing speckle compared to Fig.~\ref{fig:concept}c. This simulation shows that multisource holography can create natural defocus cues with low speckle, no ringing artifacts, and uniform energy distribution in the eyebox.

\begin{figure*}
    \centering
    \includegraphics[width=\linewidth]{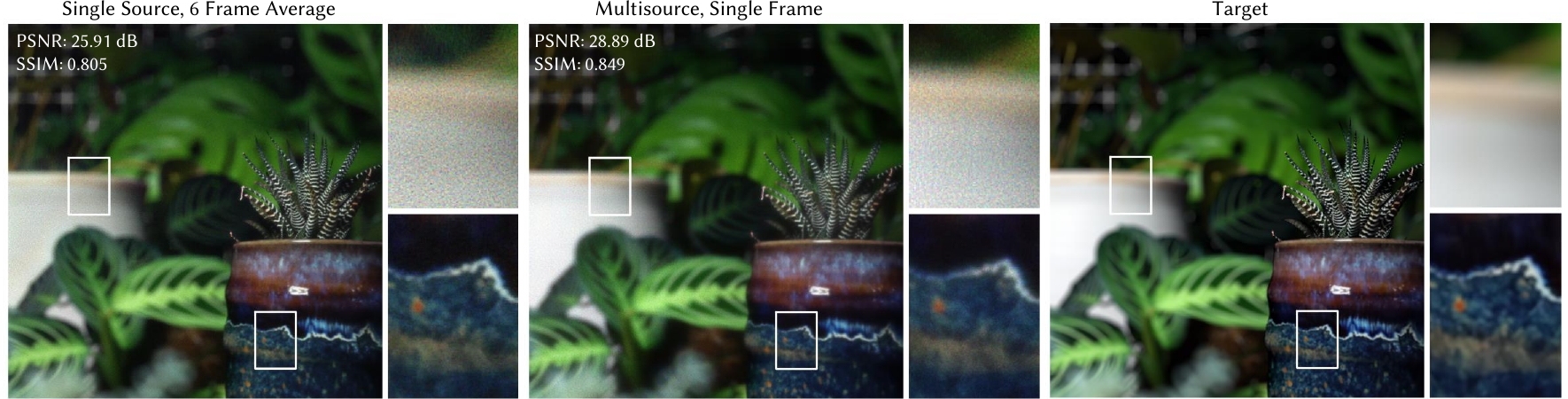}
    \caption{\textbf{Comparison with Temporal Multiplexing (Simulation):} Our multisource approach with no temporal multiplexing (one frame per color) outperforms a traditional single source hologram with a phase only SLM and 6 jointly optimized frames per color when generating a focal stack with natural blur. Our multisource simulation uses $5\times5$ sources and $\Delta m = 75$ rad/mm.
    }
    \label{fig:TM_comparion_sim}
\end{figure*}

\subsection{Source Configuration Analysis}
\label{sec:analysis}

The number of sources and their arrangement are key design choices in multisource holography, so next we analyze the impact of these parameters. Figure \ref{fig:analysis}a illustrates the effect of source spacing. Sources were arranged in a $4\times 4$ grid and the distance between neighboring sources, $\Delta m$, was varied. We simulate a $\Delta z = \SI{2}{\milli\metre}$ gap between the SLMs, and as before, we use a phase SLM as the first modulator and an amplitude SLM as the second, each with an $\SI{8}{\micro\metre}$ pixel pitch. Simulations were done for a single wavelength of $\SI{520}{\nano\metre}$, and the number of sources within the memory effect region, defined in Eq.~\ref{eq:memory_effect_condition}, are indicated by the background color in the plots.

Figure \ref{fig:analysis}a (top plot) shows PSNR as a function of $\Delta m$ for a natural scene. When the sources are within the memory effect region of the two SLMs, they create correlated patterns. Similar to the scenario with only one SLM (Sec.~\ref{sec:multisource_1slm}), the resulting output image is described by  a convolution (Eq.~\ref{eq:deconvolution_problem}). In this case, since the sources are close together, this creates a small blur instead of the dramatic ghost artifacts in Fig. \ref{fig:concept}d. Since this blur reduces noise effectively, and PSNR is not a metric that's sensitive to high resolution features, the PSNR is highest at small source spacing. However, this blur is not desirable for a high resolution holographic display.

To quantify the system's ability to display high frequency features, we simulate a binary grating with a period of two SLM pixels, the highest spatial frequency the SLM can produce. We optimize for a focal stack and measure the  Michelson contrast, $(I_{\text{max}}-I_{\text{min}})/(I_{\text{max}}+I_{\text{min}})$ in focus, averaged over a $100 \times 100$ pixel area.  Assuming an \SI{8}{\micro\metre} SLM pixel and an eyepiece with focal length $f = \SI{27.5}{\milli\metre}$, this corresponds to the contrast at 30 cycles per degree or 1 arcmin resolution, on par with the human visual system. We test with the focal plane at three different locations in the volume ($z =\SI{15.7}{\milli\metre}, \SI{20}{\milli\metre}, \SI{24.3}{\milli\metre}$) and report the average contrast.

Figure \ref{fig:analysis}a (bottom plot) shows this contrast as a function of source spacing. When $\Delta m = 0$, the sources are on top of each other. This is equivalent to a single source, which, although noisy, can display high resolution features. Once the sources move slightly apart, they are fully within the memory effect region of the two SLMs, so the the output is blurred and contrast drops. As the spacing between the sources increases, progressively more sources leave the memory effect region and the contrast at 1 arcmin increases, demonstrating that multisource holography can create high resolution features when sources are spaced sufficiently far apart. Since the memory effect cutoff (Eq.~\ref{eq:memory_effect_condition}) also depends on the gap between the SLMs, a similar trend holds when  $\Delta z$ is varied; see Supplement for an analysis of $\Delta z$. 

Next, we consider how the number of sources impacts hologram quality. Figure~\ref{fig:analysis}b shows PSNR (top) and contrast at 1 arcmin (bottom) as a function of the number of sources. Sources are arranged in an evenly spaced square grid, with $\Delta m = 50$ radians/mm spacing, which is outside the memory effect region. As the number of sources increases, there is additional despeckling due to more incoherent averaging, and this results in an increase in both PSNR and contrast at 1 arcmin (note that contrast is also negatively affected by speckle). Although there are only 2 SLMs, the image quality continues to improve far beyond two sources. This demonstrates the compressive nature of the system since it implies that each source is still able to create the correct pattern at full resolution using a limited number of degrees of freedom. 

However, compressive systems still have limits and eventually there are not sufficient degrees of freedom to uniquely create the correct content for each source. Looking at the simulated image with 100 sources, one can see haze caused by some sources creating incorrect content. Once again, PSNR does not reflect this trend, since the additional haze (which is not well captured by PSNR) is balanced by further speckle reduction. However, our contrast metric is a better proxy: around 36 sources, the contrast at 1 arcmin starts to decrease, reflecting this performance limit. This suggests that the best image quality is with a $6 \times 6$ grid, which achieves 29.4~dB PSNR, over 10 dB higher than the single source baseline.

We'd like to point out that there is a substantial design space for multisource holography. Future work includes analyzing sources that are not confined to a grid, exploring extended sources, and varying the source intensities. In addition, the source parameters could be optimized specifically for a dataset of natural images, analogous to the work of \citet{baek2021neural}. However, these explorations are out of scope for this paper.

\subsection{Time Multiplexing Comparison}

So far we have restricted our comparisons to single source holograms made with a single frame, but a common approach to speckle reduction is time multiplexing. In this approach, several holograms are displayed in rapid succession, and due to persistence of vision, the user sees an average of the displayed frames. High speed modulators have made this method increasingly practical, and prior work~\cite{choi2022time,lee2022high} has demonstrated that temporal multiplexing can create natural defocus blur with a uniform eyebox.

Our method is not meant to be a replacement to temporal multiplexing; the two approaches are orthogonal and can be combined for even more speckle reduction. Since noise reduction goes with the square root of the number of uncorrelated images, temporal multiplexing provides diminishing returns with increasing frame rate. Additional despeckling may be necessary to reduce noise to an imperceptible level, even with high speed modulators.

In addition, reducing the necessary temporal bandwidth could help with another fundamental challenge in holography: limited \'etendue, which results in a trade-off between FoV and eyebox size. One practical option to overcome this limitation is to scan the location of either the FoV or eyebox~\cite{lee2020wide}, enabling expanded \'etendue without eye tracking. However, scanning also requires temporal bandwidth, which is no longer available for despeckling. By providing substantial despeckling in a single frame, multisource holography could open new paths for increasing \'etendue.

Figure~\ref{fig:TM_comparion_sim} compares our multisource holography approach to temporal multiplexing with 6 jointly optimized frames per color. Similar to recent work using temporal multiplexing \cite{choi2021optimizing}, our holograms are computed using iterative optimization where all 6 multiplexed frames are summed together before computing the loss function. Then, all the frames are simultaneously updated by the optimizer. As in prior simulations, we target a focal stack with 15 planes and natural defocus blur. Our multisource simulation uses one phase and one amplitude SLM, 25 sources with $\Delta m = 75$~rad/mm, and only a single frame per color. 

Qualitatively, the two approaches have similar noise levels and image quality, with multisource visually outperforming temporal multiplexing in white regions. Quantitatively, multisource holography exceeds 6 frame temporal multiplexing in PSNR and structural similarity index measure (SSIM) over the focal stack. In the temporal multiplexing example, we simulated a phase only SLM, which differs from the simulations in Sec.~\ref{sec:sim_single_source} and Sec.~\ref{sec:analysis}, since this is the most realistic choice given currently available hardware. If fact, most high speed SLMs are even more restricted; the SLMs capable of this much multiplexing are typically binary or have limited bit depth, although in this simulation we assume no quantization. As the number of temporally multiplexed frames increases, the quality eventually exceeds that of multisource holography (see Supplement for an example), but it comes at the cost of temporal bandwidth.

\label{sec:results}

\section{Experimental System Calibration} \label{sec:experiment_calibration}

We have shown in simulation that multisource holography is a promising technique, but in practice, achieving high quality experimental results requires accurate knowledge of system parameters such as the source locations and positions of the two SLMs. To calibrate our experimental system, we adapt the approaches of \citet{peng2020neural} and \citet{Chakravarthula2020LearnedDisplays} to multisource holography by designing a physics-inspired forward model where unknown parameters are learned from a dataset of experimentally captured SLM-image pairs. Next, we go into the details of this model and the calibration procedure.

\subsection{System Model with Learnable Parameters} \label{sec:learnable_model}

\paragraph{SLM Model}

Our model starts with the digital values sent to the SLMs.
For each SLM, these values are passed through a learned lookup table (LUT) which describes the mapping from digital input to phase. The LUT is parameterized by 256 coefficients (one for each possible input value), and the LUT is made differentiable using 1D interpolation.

Next, the phase is convolved with a small learnable kernel that represents cross-talk between pixels due to field fringing~\cite{moser2019model, apter2004fringing, persson2012reducing}. Field fringing is a phenomenon of liquid-crystal-on-silicon (LCoS) SLMs in which the output phase is blurred by the gradual transition at pixel boundaries of the electric field that modulates the liquid crystal layer. Since this effect is sub-pixel, we upsample the phase values by $2\times$ in each direction before applying the convolution kernel ($5 \time 5$ pixels in the upsampled space).

For each SLM, the phase with field fringing is converted to an electric field (assuming uniform amplitude), yielding the complex modulation functions $\mathbf{s}_1(\vec{x})$ and $\mathbf{s}_2(\vec{x})$.

\paragraph{Source Model}

Each source is assumed to be a plane wave with learnable angle of incidence and learnable relative intensity. For each source, we parameterize the angle of incidence as a 2D location in Fourier space; simulating a delta function at that location and then taking the Fourier transform and multiplying by the relative intensity yields the input field for a given source, $\mathbf{p}(\vec{x}; \vec{m}_i)$.

\begin{figure}
    \centering
    \includegraphics[width=\linewidth]{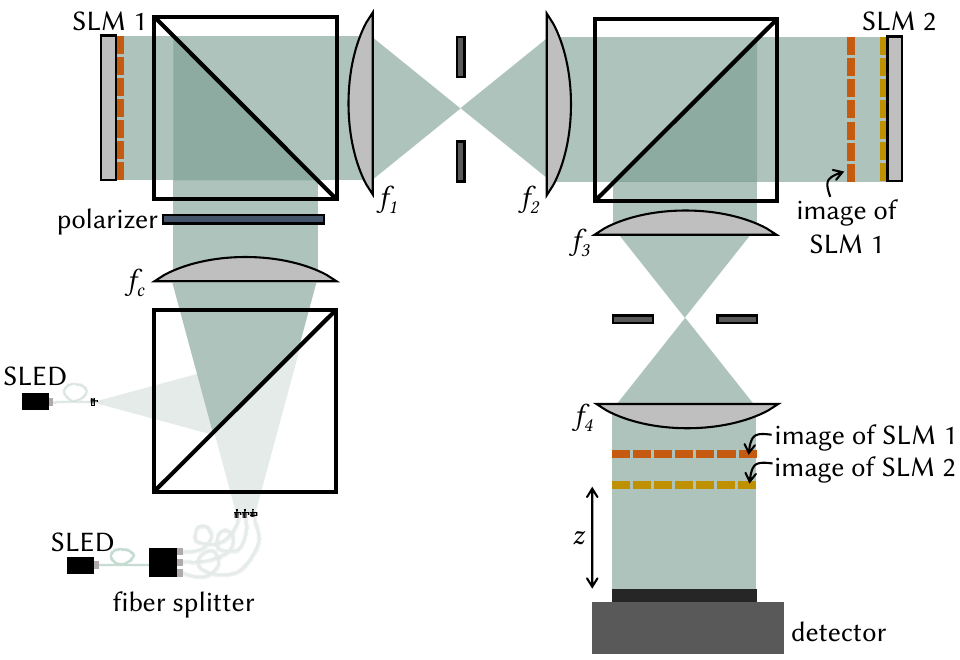}
    \caption{
       \textbf{Schematic of Experimental Setup:} Our benchtop prototype uses two SLMs with a 4$f$ system in between. A second 4$f$ relays both SLMs to the correct positions in front of a bare sensor, which is mounted on a linear motion stage. Irises in the Fourier planes remove higher orders from the SLMs. To create the multiple sources, we use a superluminescent light emitting diode (SLED) passed through a fiber splitter. Due to the low coherence of the SLED, the fiber outputs are mutually incoherent, as required by our method. A beamsplitter allows for switching between single source and multisource illumination for comparisons.
}
    \label{fig:setup_detailed_sketch}
\end{figure}

\paragraph{Propagation Model} 

We adapt the ideal ASM propagation model (Eq.~\ref{eq:ASM_propagation}) to include aberrations by multiplying the ASM kernel (Eq.~\ref{eq:ASM_kernel}) by a complex learnable pupil function. To further enable modeling of spatially varying aberrations, different locations of the input field should have variable pupil functions. Therefore, we learn a $9 \times 16$ grid of pupil functions, and we perform bilinear interpolation to get the intermediate values.

However, applying a fully-spatially varying model is very computationally intensive. To avoid computing a different pupil function for each point of the input field, we instead take a stochastic, patch-based approach: during optimization, we randomly choose a patch of the input field (about $1200 \times 1200$ pixels in the upsampled coordinates) and use the pupil function that corresponds to the center of that patch. Over the course of optimization, this approximates the smoothly varying aberrations, with the added advantage of reducing the memory requirements of the model by only simulating a fraction of the FoV in each iteration. See Supplement for more details on how aberrations are parametrized.

\paragraph{SLM Alignment} 

If the two SLMs are not perfectly aligned with sub-pixel accuracy, we need to account for their relative positions in the model. After propagating the field from the first SLM, we apply a learned warping function that transforms the field into the coordinate space of the second SLM. Our warping function, based on the thin-plate spline model (TPS) of \citet{duchon1977splines}, can account for non-radial distortion between the two SLMs, enabling accurate alignment even when there are non-ideal optics between the modulators. The warping is implemented in a differentiable manner in Kornia~\cite{riba2020kornia} using bilinear interpolation separately on the real and imaginary parts of the complex field.

\paragraph{Model Summary} 

We put together all the components of the model as follows: starting with the first SLM, we use our SLM model to covert the digital input values into a complex modulation function. This is multiplied by the source field, then propagated a distance $\Delta z$ using our modified ASM propagator with spatially varying pupil functions. The field is then warped to match the coordinate space of the second SLM and multiplied by the complex modulation function $\mathbf{s}_2(\vec{x})$, which, once again, is computed with the SLM model described above. Finally, we apply the ASM propagator with spatial variance a second time to propagate a distance $z$, then take the absolute value squared to simulate intensity on the sensor. This process is repeated for each source in the system while summing the contributions.

\subsection{Calibration Procedure} \label{sec:calibration_procedure}
To fit the learnable parameters of our model, we collect an experimental dataset of SLM-image pairs and optimize for the unknown parameters using gradient descent in PyTorch. We use random patterns on the SLM, which have similar statistics to the random phase holograms we aim to display. To further facilitate the optimization process, we apply a Gaussian filter on the input phase with a standard deviation varying from 4 pixels to zero pixels (no blur). This creates training data with larger features that are especially helpful when optimizing the TPS and the source position parameters, which do not converge correctly with high-frequency content alone. We capture datasets with both single source illumination and multisource illumination.

Since the low frequency SLM inputs are less sensitive to field fringing and aberrations, we use the single source blurred patterns to optimize the TPS warping function before fitting the rest of the model. We also optimize a second similar warping function to align the final intensity to the camera capture. Once the alignment functions are close to accurate, we use the remaining single source dataset with all spatial frequencies to fit the other parameters.

After the single source model is optimized, we use the multisource dataset to fit the source locations and intensities. Finally we fine-tune the other parameters using the multisource data to get the complete model. We repeat this process for each color separately.

Note that unlike many learned models in prior work \cite{Choi2021NeuralDisplays, Chakravarthula2020LearnedDisplays}, our model does not contain any black-box neural networks; all parameters are physically meaningful. This limits the number of learnable parameters, which in turn means less training data is required, the model optimizes quickly, and the chance of over-fitting is low. For example, our training dataset contains only about 300 captures per color per source configuration, and training takes approximately 10 minutes on an Nvidia GV100. Although we only capture training data at a single propagation distance $z$, we find that the model extends well to other planes without retraining.

\subsection{Active Camera-in-the-Loop} \label{sec:ACITL}

To highlight the potential of multisource holography, we additionally use the ``active'' camera-in-the-loop (CiTL) method proposed by \citet{peng2020neural}, where feedback from a camera in the system is used to fine-tune the SLM pattern(s) for a specific image or focal stack. We pre-optimize the SLM patterns using our learned model, display the patterns on the experimental system, and continue optimization while replacing the model output with the captured image before back-propagation. For focal stacks, we capture the experimental images at a different location in the volume at each iteration, and we fine-tune the alignment between the capture to the model output using cross-correlation on a patch-wise basis. Final results are captured after updates are complete, with one static pair of SLM patterns for all depths.

\begin{figure}
    \centering
    \includegraphics[width=\linewidth]{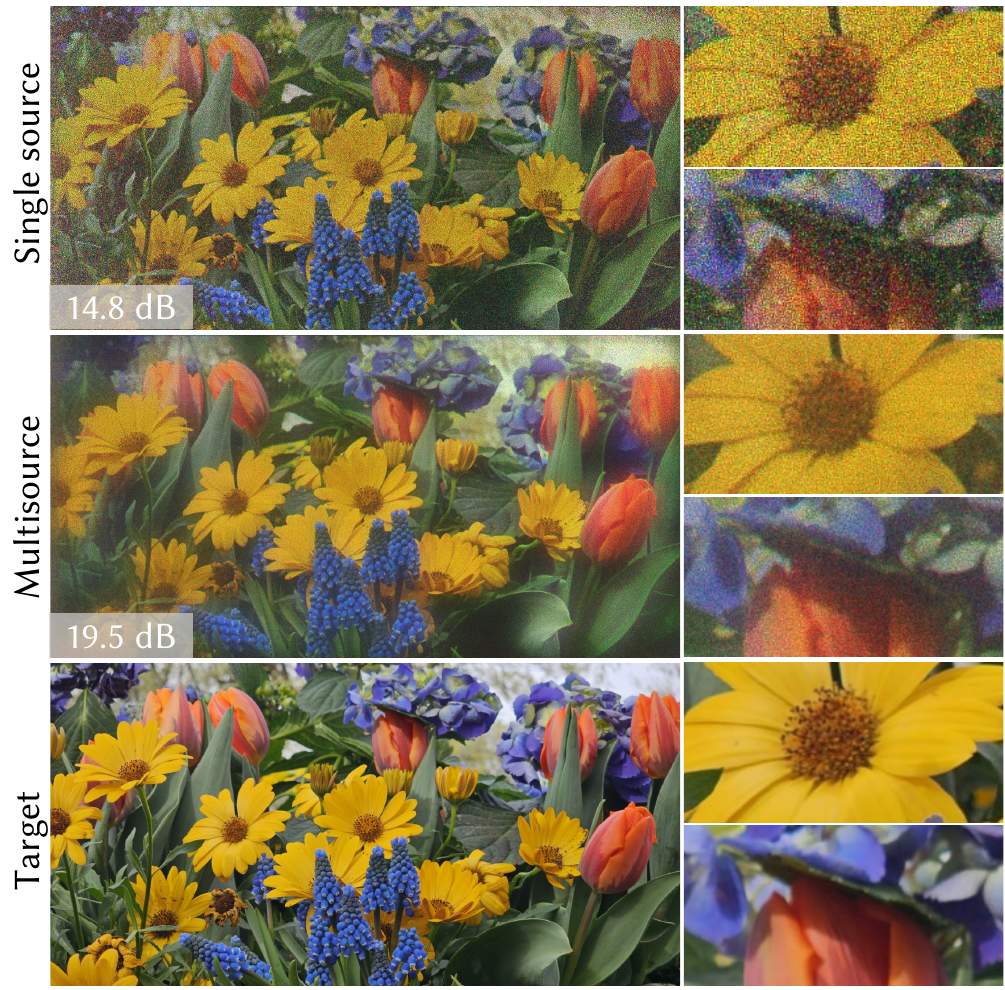}
    \caption{\textbf{2D Results (Experiment):} Although a single source random phase hologram can theoretically control speckle well for a 2D image, the experimental 2D capture (top) has visible speckle when one zooms in. Our multisource configuration with $4\times 4$ sources (middle) has noticeably reduced speckle while maintaining high frequency features. PSNR is shown in the bottom left.
    }
    \label{fig:results:flowers_2d}
\end{figure}

\begin{figure*}
    \centering
    \includegraphics[width=\textwidth]{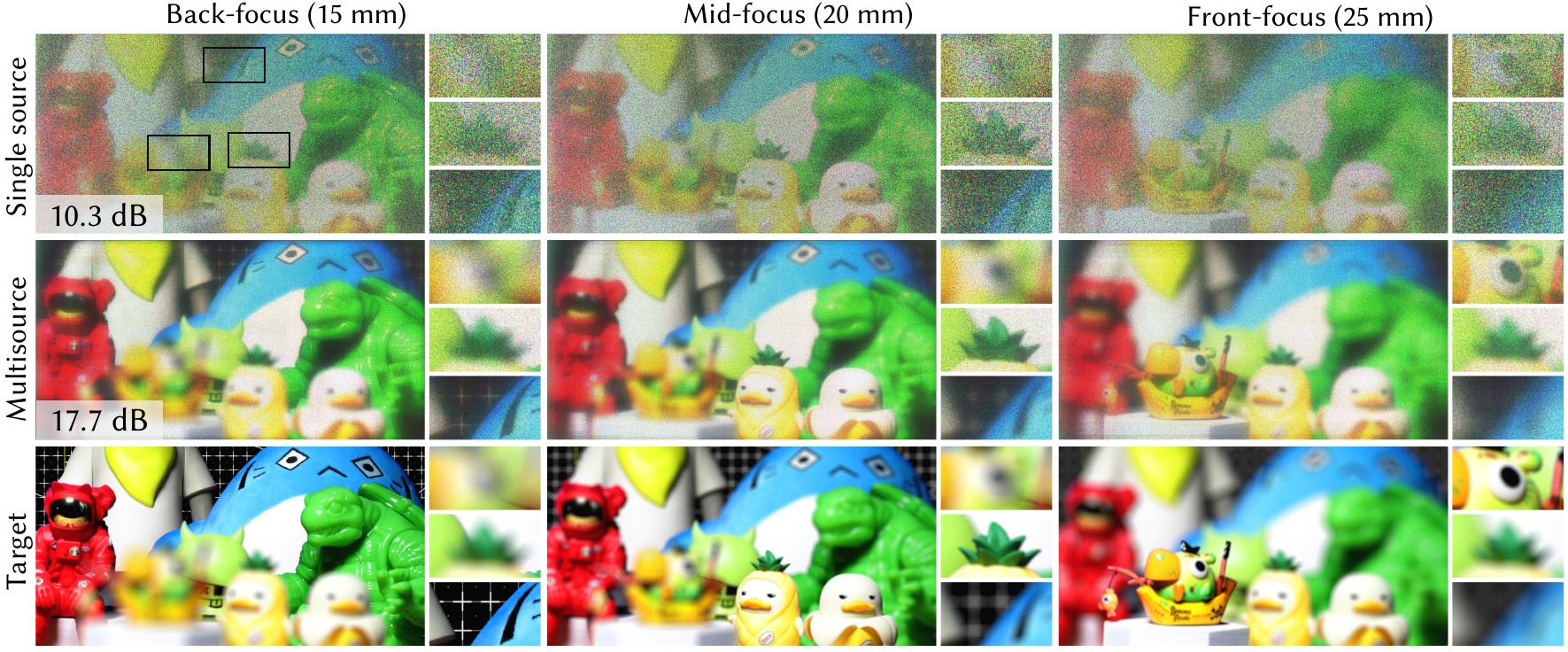}
    \caption{
       \textbf{Focal Stack Results (Experiment)}:
       Focal stacks created by a single source hologram with random phase (top) suffer from severe speckle noise since there are insufficient degrees of freedom on the SLM to control speckle throughout a 3D volume. Our multisource approach with $4\times4$ sources (middle) greatly reduces speckle, enabling experimental focal stacks with natural defocus cues. PSNR calculated over the full focal stack is shown in the bottom left.
}
    \label{fig:experiment_focalstack}
\end{figure*}

\section{Experimental Results}

We demonstrate multisource holography on a benchtop experimental system, depicted in Fig.~\ref{fig:setup_detailed_sketch}. To create the multiple sources, we split the output of a fiber-coupled light source using cascaded 1:4 fiber splitters (Thorlabs TWQ560HA) to create 16 different sources, which are arranged in a $4 \times 4$ grid. By choosing a superluminescent light emitting diode (SLED, Exalos EXC250011-00), which has a very short coherence length, we find that the outputs of the 16 different fibers are mutually incoherent without explicitly adding path length differences. However, the SLED has a spectral bandwidth of about \SI{10}{\nano\metre}, which is not accounted for in our model, and we discuss this limitation more in Sec.~\ref{sec:discussion}. Although the spectral bandwidth of a laser would match our model better and result in improved resolution \cite{deng2017coherence}, we found that the longer coherence length of a laser made it challenging to consistently break the coherence between fiber outputs, even with added path length differences. This is not a fundamental challenge as one could use an array of laser diodes instead of splitting a single laser output.

The multiple sources are spaced $\SI{4}{\milli\metre}$ apart, held in a 3D printed housing. Combined with the $f_c = \SI{500}{\milli\metre}$ collimating lens, this yields $\Delta m = 79$ rad/mm, $99$ rad/mm, and $110$ rad/mm for red, green, and blue respectively, which are outside the memory effect region of Eq.~\ref{eq:memory_effect_condition}. The angles of incidence at the SLM are within $\pm 0.69^\circ$ (see Eq.~\ref{eq:angle_of_incidence}), which is well within the paraxial approximation as assumed in Sec.~\ref{sec:methods}. A beamsplitter in front of the sources lets the illumination be toggled between the multisource configuration and a traditional single source, and a linear polarizer  ensures the beam is correctly polarized for the SLMs.

The system uses two phase only LCoS SLMs (Holoeye Pluto-2.1-VIS-016), and a 4$f$ system with 1:1 magnification ($f_1=f_2=\SI{200}{\milli\metre}$) relays the first SLM to a distance $\Delta z = \SI{2}{\milli\metre}$ behind the second SLM. A second 4$f$ system ($f_3=\SI{200}{\milli\metre}$, $f_4=\SI{150}{\milli\metre}$) relays the SLMs to the camera sensor. Irises in the Fourier planes of both 4$f$ systems filter higher orders from the SLMs.

SLM patterns are optimized using the calibrated model outlined in Sec.~\ref{sec:experiment_calibration}. SLMs are initialized with uniform random phase, and we jointly optimize both SLMs, even for the single source case. Different SLM patterns are optimized for each color (central wavelengths at \SI{638}{\nano\metre}, \SI{510}{\nano\metre}, and \SI{455}{\nano\metre}, for red, green, and blue respectively) and displayed in sequence.

Images are captured on a monochrome camera sensor (XIMEA MC089MG-SY), which is mounted on a brushless translation stage (Thorlabs DDS050) to enable focal-stack capture from $z = \SI{15}{\milli\metre}$ to $z = \SI{25}{\milli\metre}$, defined in SLM space. Note that the actual distances at the camera are slightly less due to the demagnification of the second 4$f$ system. Since the sensor is monochrome, color results are captured sequentially and combined in post-processing. After capture, images are rectified into the SLM coordinate space using bilinear interpolation, un-modulated areas of the image are cropped out, and the relative intensities of the color channels are adjusted.

\subsection{2D Results}

Figure~\ref{fig:results:flowers_2d} shows a 2D experimental capture on our system at $z = \SI{20}{\milli\metre}$ comparing single source and multisource holograms. Although a traditional single source holographic display can theoretically create very high quality 2D holograms, in practice there is still speckle noise visible in a random phase hologram. Even in the 2D scenario where single source performs quite well, multisource holography still provides noticeable despeckling, improving PSNR by 4.7 dB in this example. Both results are optimized with active CiTL as described in Sec.~\ref{sec:ACITL}; see Supplement for results without this fine tuning.

\subsection{Focal Stack Results}

However, the true benefits of multisource holography become most apparent when displaying 3D content. Using our calibrated model, we optimize the SLM patterns while targeting a focal stack with natural blur. We use the same blur parameters and propagation distances as the simulations (Sec.~\ref{sec:simulation}).

Figure~\ref{fig:experiment_focalstack} shows the experimentally captured results. As expected from our simulations, the hologram made with a single source is severely corrupted by speckle. In comparison, multisource holography can generate low speckle images over the whole volume, complete with natural defocus cues, resulting in a 7.4 dB PSNR increase calculated on the full focal stack. As a reminder, our multisource holograms are random phase, which creates an approximately uniform energy distribution in the eyebox (see Supplement for a visualization), and are produced with only one frame per color. Similar to the 2D images, these results are all captured with active CiTL; versions without active CiTL are included in the Supplement.

\section{Discussion}
\label{sec:discussion}
We have demonstrated both in simulation and experiment that multisource holography can provide significant despeckling without resolution loss, enabling focal stacks with realistic blur. However, there are several directions for further investigation.

\paragraph{Pupil-Aware Holography}
Our holograms (like most in the literature) are simulated assuming the entire eyebox is fully contained within the user's pupil. This is atypical for conventional (non-holographic) near-eye displays where the eyebox is usually larger than the pupil size to give users freedom to move their eyes without leaving the eyebox. However, in a holographic display, substantial artifacts can occur when only a fraction of the eyebox enters the user's eye. Our initial simulations suggest that multisource holography could improve image quality given unknown pupil locations.

\begin{figure}
    \centering
    \includegraphics[width=\linewidth]{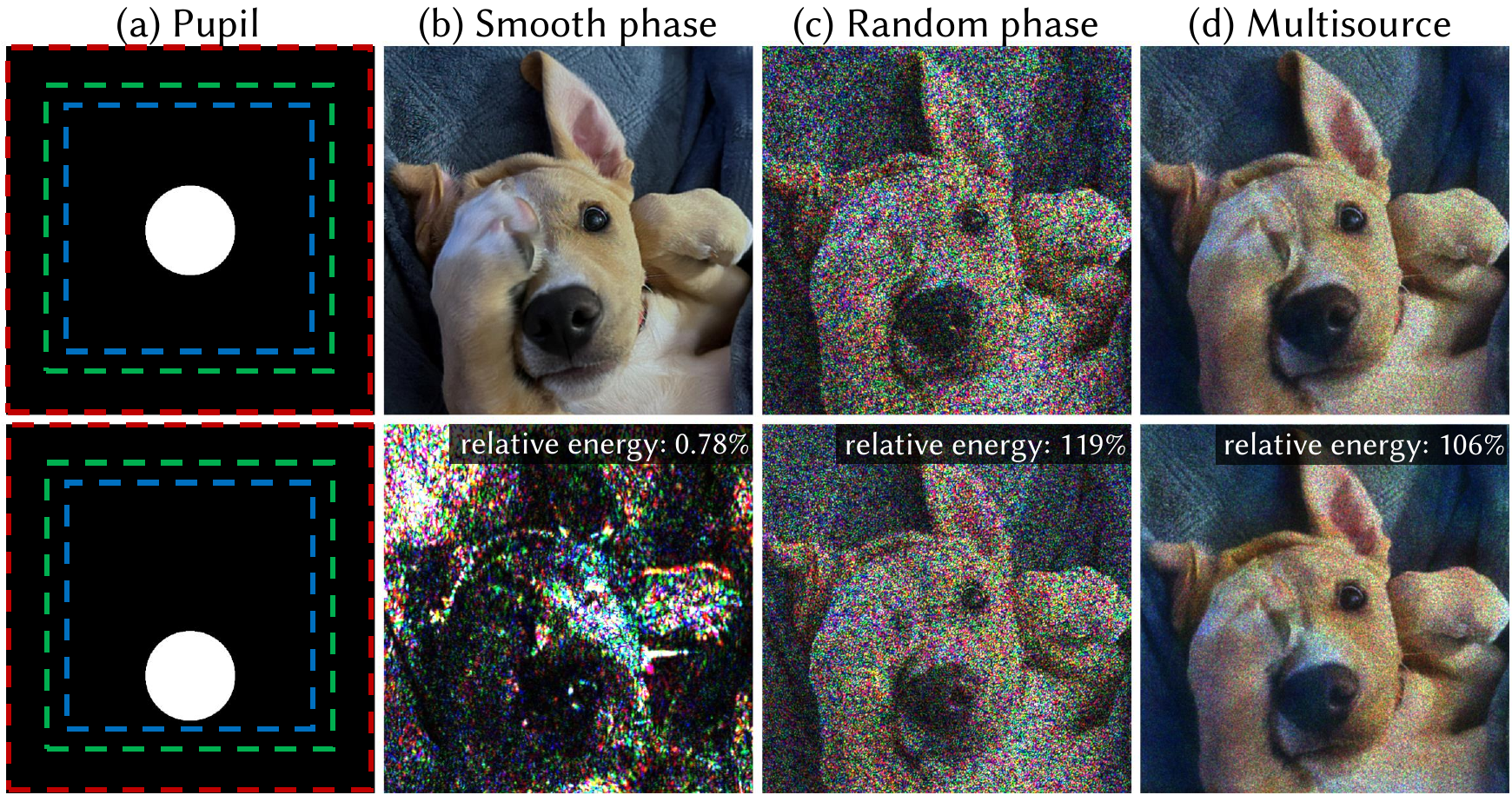}
    \caption{\textbf{Pupil-Invariance (Simulation):}
    When the user's pupil does not cover the full eyebox, holograms can have significant artifacts, even for 2D images.
    To demonstrate, we optimize holograms with the pupil aware loss of \citet{chakravarthula2022pupil}. We show examples at two different pupil positions in the eyebox, visualized in (a), where the eyebox extent for each color is depicted with dotted lines. The total intensity of the simulated image relative to a centered pupil is shown in the top right of each simulation. Smooth phase holograms (b) can create high quality images when the pupil is centered, but when the pupil is translated, image content is highly corrupted and has very low intensity. Random phase holograms (c) have approximately uniform intensity when the pupil moves but the image is very noisy due to speckle. Multisource holography (d) can create a low noise image that's invariant to the pupil position, which is desirable for a practical display.
    }
    \label{fig:pupil_invariance}
\end{figure}

To demonstrate, we simulate 2D holograms optimized using the pupil-aware loss proposed by \citet{chakravarthula2022pupil}, in which random pupil locations are sampled during optimization (Fig.~\ref{fig:pupil_invariance}). Smooth phase holograms (Fig.~\ref{fig:pupil_invariance}b) have excellent image quality when the pupil is centered, but a pupil at the edge of the eyebox sees a low intensity, completely incorrect version of the image. Random phase holograms (Fig.~\ref{fig:pupil_invariance}c) have approximately uniform intensity but are corrupted by speckle regardless of pupil position, even for a 2D images with no focal cues. In comparison, multisource holography (Fig.~\ref{fig:pupil_invariance}d) can produce a clean image for pupil locations over the whole eyebox. Extending this concept to light fields~\cite{choi2022time, padmanaban2019holographic} is another direction of future work.

\paragraph{Source Design and \'Etendue Expansion}
Although we analyzed several important parameters of the source design in Sec.~\ref{sec:analysis}, we restricted our analysis to a grid of uniform intensity sources within the \'etendue of the native SLM. There may be additional performance gains from different source configurations such as extended sources, optimized source locations, or variable source intensities. In addition, by increasing the spacing between sources, multisource holography may be able to expand system \'etendue, similar to the work of \citet{jo2022multi}, helping with another fundamental problem in holographic displays.

\paragraph{Multisource Holography with 1 SLM}
Using two SLMs may not be feasible for all applications. An alternative is to replace one of the SLMs with a static DOE. This creates an angularly-selective response similar to the two SLMs, breaking correlations between sources and enabling many of the benefits of multisource holography. However, the reduced degrees of freedom mean that fewer sources can simultaneously generate the correct pattern, so the amount of despeckling will be reduced. To improve performance with only a single active modulator, co-optimization of the DOE with the other system parameters could be investigated, similar to the work of~\citet{baek2021neural}.

\paragraph{Compact Architecture}
Our experimental system is a large benchtop setup containing multiple 4$f$ systems, but we envision multisource holography could be built into a compact architecture. Starting with the design of \citet{kim2022holographic}, which uses a waveguide to illuminate a reflective phase only SLM, we propose coupling the multiple sources into the waveguide to generate the multisource illumination. For the second SLM, we suggest using a transmissive amplitude modulator, placed just before the eyepiece. However, without a 4$f$ relay system, SLM higher orders must be taken into account in the model~\cite{gopakumar2021unfiltered} or filtered using compact volume holograms~\cite{bang2019compact}.

\paragraph{SLED Bandwidth}
We took advantage of the short coherence length of the SLED to create the multiple sources used in our experimental setup. However, the SLED has a bandwidth of about \SI{10}{\nano\metre} while our model and analysis in Sec. \ref{sec:analysis} assumes monochromatic light. We include in the Supplement a complete model that accounts for the spectral bandwidth of the source and a practical optimization strategy based on \citet{peng2021speckle} for this scenario. However, modeling a larger bandwidth source has higher computational cost, so we chose to assume monochromatic illumination during optimization. We expect our results would improve with more accurate modeling of the SLED, but we found that our monochromatic model was sufficient to show the benefits of multisource holography. See Supplement for a visualization of the effect of the SLED.

\paragraph{Computation Speed}
Computational cost is a limitation of our method, since the image formation model requires separately simulating the contributions from each source. Furthermore, all our simulations were conducted at $2\times$ resolution in each dimension, resulting in computation times of about half an hour to generate a focal stack. For example, for a 1080 $\times$ 1920 modulator with 16 sources, we run 2000 iterations, each of which takes about 0.8 sec. Upsampling may not be necessary in all scenarios, and in these cases computation time drops to about 0.2 sec per iteration, but compute is still a limitation. To address this, neural networks offer a promising path towards real-time computation, as they have already been demonstrated for single source holography ~\cite{peng2020neural, shi2021towards,yang2022diffraction}, albeit only for smooth phase so far. Adapting these approaches to multisource holography will be necessary for a practical display.

\section{Conclusion}
\label{sec:conclusion}

We introduced a new architecture for holographic displays that uses an array of mutually incoherent sources and two SLMs to reduce speckle. To our knowledge, our design is the first single-frame method that can generate low speckle holograms at full resolution with realistic focal cues and a uniform eyebox. We analyzed the concept in simulation, explored the design space, and validated with a benchtop experimental setup capable of producing high quality focal stacks. In conclusion, we believe multisource holography is a promising path to address some of the key open problems in holographic displays.

%%%%%%%%%%%%%%%%%%%%%%%%%%%%%%%%%%%%%%%%%%%%%%%%%%%%%%%%%%%%%%%%%%%%%%%%%%%%%%%%%%%%%%%%

% Bibliography
\bibliographystyle{ACM-Reference-Format}
\bibliography{references}

\end{document}